\documentclass[aps,prl,twocolumn,superscriptaddress]{revtex4}
\usepackage{mathrsfs}
\usepackage{epsfig}
\usepackage{graphicx}
\usepackage{amsfonts}
\usepackage[figuresright]{rotating}
\usepackage{amssymb}
\usepackage{amsmath}
\usepackage{dcolumn}
\usepackage{bm}
\usepackage{color}

\def\be{\begin{equation}} \def\ee{\end{equation}}
\def\bea{\begin{eqnarray}} \def\eea{\end{eqnarray}}

\newcommand{\WQCASQC} {Wilczek Quantum Center and Key Laboratory of Artificial Structures and Quantum Control, School of Physics and Astronomy, Shanghai Jiao Tong University, Shanghai 200240, China}

\newcommand{\SRCQC}{Shanghai Research Center for Quantum Sciences, Shanghai 201315, China}

\begin{document}
\title{Dynamical transitions and critical behavior between discrete time crystal phases}
\author{Xiaoqin Yang}
\affiliation{\WQCASQC}

\author{Zi Cai}
\email{zcai@sjtu.edu.cn}
\affiliation{\WQCASQC}
\affiliation{\SRCQC}

\begin{abstract}  
In equilibrium physics, spontaneous symmetry breaking and elementary excitation are two concepts closely related with each other: the symmetry and its spontaneous breaking not only control the dynamics and spectrum of elementary excitations, but also determine their underlying structures. In this paper, based on an exactly solvable model, we propose a phase ramping protocol to study an excitation-like behavior of a non-equilibrium quantum matter: a discrete time crystal phase with spontaneous temporal translational symmetry breaking. It is shown that slow ramping could induce a dynamical transition between two $Z_2$ symmetry breaking time crystal phases in time domain, which can be considered as a temporal analogue of the soliton excitation spacially sandwiched by two degenerate charge density wave states in polyacetylene. By tuning the ramping rate, we observe a critical value at which point the transition duration diverges, resembling the critical slowing down phenomenon in nonequilibrium statistic physics. We also discuss the effect of stochastic sequences of such phase ramping processes and its implication to the stability of the discrete time crystal phase against noisy perturbations.

\end{abstract}


\maketitle

Usually, the ground state energy of an equilibrium system does not have much to do with its observable behavior.  What's  physically important are the properties of low-lying excited states, which are likely to be excited owing to weak external fields or relatively low temperatures. For example, the thermal and elastic properties of a solid are determined by a few number of lattice wave excitations  known as  phonons\cite{Anderson1997}. Non-equilibrium quantum matter fundamentally differs from its equilibrium counterparts, and  has received considerable interest in various fields ranging from ultracold atoms\cite{Eisert2015}  to solid state physics\cite{Mankowsky2014} over the past decade. However, compared to equilibrium cases, much less is known about the ``excitation'' of non-equilibrium quantum matter, even its definition may be questionable, let alone its relationship with fundamental properties of non-equilibrium quantum matter, {\it e.g.} the symmetries and their spontaneous breaking.

As a prototypical example of  nonequilibrium quantum matter, the time crystal (TC) phase has allowed new possibilities for the  spontaneous symmetry breaking (SSB) paradigm\cite{Wilczek2012}, and have attracted considerable interest in its different forms\cite{Shapere2012,Li2012,Wilczek2013,Sacha2015,Else2016,Khemani2016,Yao2017,Russomanno2017,Gong2018,Huang2018,Iemini2018,Das2018,Zhu2019,Kozin2019,Khasseh2019,Cai2020,Chinzei2020,Yao2020}. Such a intriguing state, despite being proven to be forbidden in equilibrium~\cite{Bruno2013,Watanabe2015}, has been experimentally realized in non-equilibrium settings with periodic driving~\cite{Choi2017,Zhang2017}. Its physical observables develop persistent oscillations whose periods are an integer multiple of the Hamiltonian period, thus spontaneously break the discrete temporal translational symmetry(DTTS). In equilibrium systems, the SSB is closely related to the elementary excitation: it does not only affect the dynamics and spectrum of an elementary excitation, but also determines its structure. For example, for a one-dimensional (1D) system, a spontaneous discrete spatial translational ({\it e.g.} $Z_2$) symmetry   breaking  allows certain  soliton-like excitation:  a topological defect sandwiched by two $Z_2$ symmetry breaking phases\cite{Su1979}. A profound  question is how to generalize such an idea into the non-equilibirum quantum matters with intriguing SSB absent in equilibrium physics ({\it e.g.} DTTS breaking).

\begin{figure}[htb]
\includegraphics[width=0.99\linewidth]{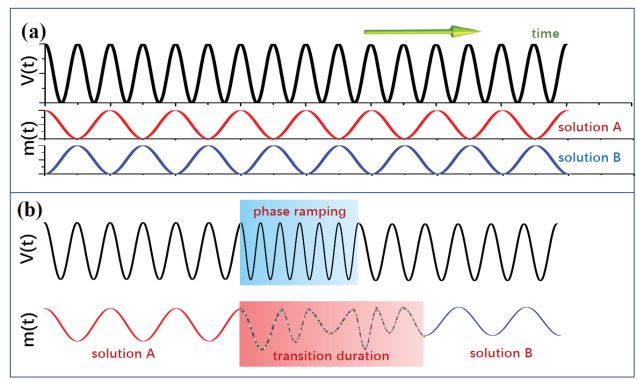}
\caption{(Color online).
Schematic diagram of (a) a period doubling dynamics in the presence of periodical driving and two degenerate TC phases and (b)the phase ramping protocol in our model and the dynamical transition induced by it.} \label{fig:fig1}
\end{figure}

In this paper, we address this issue by studying  a driven 1D interacting bosonic model that can manifest a sub-harmonic response for  physical quantities, a signature of a discrete time crystal (DTC) phase.  We impose a time-dependent perturbation on top of the periodical driving, which transiently breaks the original time translational symmetry, and we then monitor the response of the physical observable. It is shown that a slow perturbation may induce a dynamical transition between two $Z_2$ symmetry breaking DTC phases, as shown in Fig.\ref{fig:fig1}.  By tuning the ramping velocity, one can observe a critical point, at which the transition duration diverges, a reminiscence of the critical slowing down phenomenon\cite{Hohenberg1977,Taeuber2014}. Finally, we discuss the effect of multiple random phase ramping processes and its implication to the stability of the discrete time crystal phase against noisy perturbations.

{\it Model and method:} We consider a 1D hard-core bosonic model with an infinite long-range interaction. The Hamiltonian reads as follows:
\begin{equation}
H=-J\sum_i (\hat{b}_i^\dag \hat{b}_{i+1}+h.c)-\frac{V(t)}L \sum_{ij}(-1)^{i-j} \hat{n}_i \hat{n}_j \label{eq:Ham}
\end{equation}
where $J$ is the single-particle nearest-neighbor(NN) hopping amplitude, $\hat{n}_i=\hat{b}_i^\dag \hat{b}_i$ is the particle number operator at site $i$. $V(t)$ is the strength of the all-to-all interaction, which is time-dependent but does not decay with distance. $L$ is the system size of the 1D lattice, and the prefactor $\frac 1L$ in front of the interaction terms of Eq.(\ref{eq:Ham}) guarantees that the total interacting energy linearly scales with the system size.  In a bipartite lattice ({\it e.g,} a 1D lattice as in our case),  Eq.(\ref{eq:Ham}) indicates that the interaction between a pair of bosons are attractive (repulsive)  if they are located in the same (different) sublattice. Such an infinitely long-range interaction has been realized in recent high-finesse cavity experiments\cite{Landig2016,Hruby2018} by coupling  bosons to the cavity vacuum mode whose period is twice of that of the optical lattice. The total particle number $N$ is conserved in our system, and  we focus on the case of half-filling ($N=L/2$) throughout this paper. In the equilibrium case ($V(t)=V_0$), the ground state of the 1D  Ham.(\ref{eq:Ham}) is always a Mott-insulator with a charge-density-wave (CDW) order for arbitrary positive $V_0$.

Owing to the all-to-all coupling of the interaction in Ham.(\ref{eq:Ham}), in the thermodynamical limit the mean-field method is not an approximation but an exact method. In particular,  it provides an exact description of both the equilibrium properties and real-time dynamics of the system\cite{Blass2018,Igloi2018} (see also Supplementary Material(SM)\cite{Supplementary}).  The mean-field method allows us to decouple the all-to-all interaction  by introducing an auxiliary staggered field which is self-consistently determined during the evolution. The Ham.(\ref{eq:Ham}) can be expressed as:
\begin{equation}
\bar{H}(t)=-J\sum_i [b_i^\dag b_{i+1}+h.c]+ m(t)V(t)\sum_i (-1)^i \hat{n}_i \label{eq:MF}
\end{equation}
where $m(t)=\langle\Psi(t)|\frac 1L(-1)^i\hat{n_i}|\Psi(t)\rangle$ and $|\Psi(t)\rangle$ is the wavefunction of the system at time $t$. The time evolution under the Ham.(\ref{eq:MF}) can be solved exactly by performing the Jordan-Wigner transformation to transform 1D hard-core bosons into spinless fermions, and the Ham.(\ref{eq:MF}) transform into a non-interacting fermionic model.

We choose the periodic boundary condition (PBC), which allows us to perform the Fourier transformation, after which the fermionic Hamiltonian turns to
\begin{equation}
\bar{H}(t)=\sum_k
  \begin{array}{cc}
   [c_k^\dag & c_{k+\pi}^\dag]\\
    &
\end{array}
 \left[
  \begin{array}{cc}
    \varepsilon_k & m(t)   \\
  m(t) &\varepsilon_{k+\pi}
\end{array}
\right]
\left[
  \begin{array}{c}
    c_k  \\
    c_{k+\pi} \\
\end{array}
\right]\label{eq:Hamk}
\end{equation}
where the summation is over the momentum in the first Brillouin zone of Ham.(\ref{eq:MF}) ($k\in [-\frac\pi2,\frac\pi2]$), and $c_k$ ($c^\dag_k$) denotes the annihilation(creation) operator of the spinless fermion. $\varepsilon_k=-2J\cos k$, thus $\varepsilon_k=-\varepsilon_{k+\pi}$. Eq.(\ref{eq:Hamk}) indicates that the dynamics of the system can be considered as a collective behavior of different k modes, each of which is a two-level quantum system subjected to time-dependent field that was  self-consistently  determined as  $m(t)=\frac 1L\sum_k \langle \Psi_k(t)|c^\dag_k c_{k+\pi}|\Psi_k(t)\rangle$.

{\it Discrete time crystal.} Despite the triviality of the ground state phase diagram, the system can exhibit rich dynamical behavior in the presence of a time-dependent $V(t)$. For instance, in a quantum quench protocol, one can start from a ground state of Ham.(\ref{eq:Ham}) with $V(t=0)=V_i$, suddenly change it to a different value $V(t>0)=V_f$ and let the system evolve under this new Hamiltonian. It has been shown that the long-time dynamics of this model can exhibit either persistent oscillations or thermalization depending on the choice of initial states\cite{Chen2020} (see SM\cite{Supplementary}). The dynamical behavior is even richer and more interesting when we introduce periodical driving into Ham.(\ref{eq:Ham}), e.g. $V(t>0)=V_f+\delta \cos2\pi t$ with $\delta$ the driving amplitude. We observe that\cite{Supplementary} depending on the different choices of the initial states  and  driving amplitude $\delta$, the long-time dynamics could exhibit a periodic oscillation with a frequency that is either identical to or independent of the driving frequency; the former can be considered as a synchronization phenomenon which has been observed in periodically-driven integrable systems\cite{Russomanno2012}. In addition, it can also exhibit  quasi-periodic oscillations with a multi-period structure\cite{Supplementary}.

 \begin{figure}[htb]
\includegraphics[width=0.99\linewidth,bb=58 105 780 769]{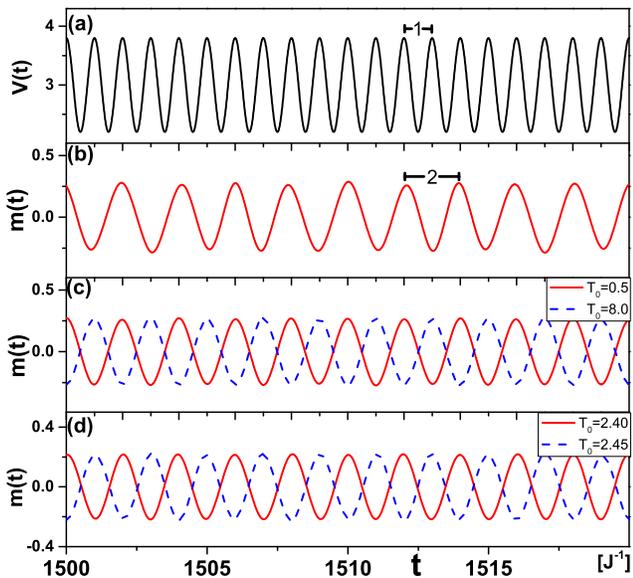}
\caption{(Color online).
(a)Periodical driving $V(t>0)=V_f+\delta \cos2\pi t$ with a period $T_0=1$. (b) Time crystal dynamics with a period $2$ in the absence of ramping. (c) Long-time dynamics of $m(t)$ with slow and fast ramping, which correspond to two  ``degenerate'' TC phases.  (d)Long-time dynamics of $m(t)$ with an intermediate ramping rate close to the dynamical transition point.  The parameters are chosen as $V_i=10J$, $V_f=3J$, $\delta=0.8J$ and $L=5000$.   } \label{fig:fig2}
\end{figure}

 \begin{figure}[htb]
\includegraphics[width=0.99\linewidth,bb=70 76 780 683]{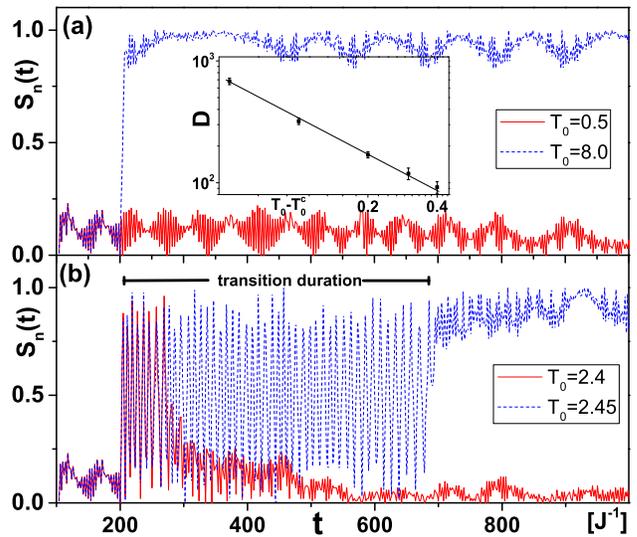}
\caption{(Color online).
Dynamics of the relative displacement of the peak positions $S_n(t)$ with (a) fast and slow  and (b) intermediate ramping rates close to the dynamical transition point $T_0^c=2.41$. The inset of Fig.\ref{fig:fig2} (a) shows the transition duration close to the dynamical critical point.   Other parameters are chosen to be the same as those in Fig.\ref{fig:fig2}} \label{fig:fig3}
\end{figure}

Most interesting dynamics can be observed in the intermediate driving regime (see Fig.\ref{fig:fig2} b), where  $m(t)$ exhibits a persistent oscillation whose period is twice that of  the external driving period: a signature of DTC\cite{Else2019} that the DTTS in Ham.(\ref{eq:Ham}) $H(t)=H(t+T)$ has been spontaneously broken ($m(t)=m(t+2T)\neq m(t+T)$ with $T=1$ is the period of driving). This phenomena is rooted in the non-linearity of the self-consistent mean-field equation of motion. However, unlike the period doubling phenomena in the non-linear classical systems ({\it e.g.}a driven-dissipative pendulum\cite{Thompson2002}) or the ``dissipative TC'' in open quantum systems\cite{Gong2018},  our system is dissipationless without entropy generation. Ham.(\ref{eq:Ham}) is free from disorder, hence it is the integrability\cite{Yuzbashyan2005} rather than the many-body localization\cite{Else2016} that prevents our model from being heated to an infinite temperature state.

{\it Dynamical transition between $Z_2$ symmetry breaking DTC phases:} Despite the richness of the dynamics behavior, here we will study neither the global non-equilibrium phase diagram of our model, nor the mathematical origin behind this DTC phase. Instead, we will use this exactly solvable model as a starting point to study a non-trivial dynamical behavior of the DTC phase, which can be considered as a temporal analogy of the soliton excitation in  Su-Schrieffer-Heeger (SSH) model\cite{Su1979}.  Owing to the spontaneous breaking of the DTTS, the DTC phase are supposed to be two-fold  ``degenerate'', each of which has a period $2$ and can be connected to the other one by shifting a half-period (1) along the temporal direction. Motivated by the soliton excitation 1D CDW system  which separates these two  $Z_2$ symmetry breaking ``degenerate'' states in space\cite{Su1979}, to realize such an object, one needs to impose a temporal perturbation on the system that transiently breaks the original time translational symmetry.

Here, we introduce an additional linear ramping of the phase on top of the periodic driving:
\begin{equation}
V(t)=V_f+\delta \cos2\pi[t+\phi(t)], \label{eq:ramping}
\end{equation}
where $\phi(t)=\frac{t-t_i}{T_0}$ for $t\in [t_i,t_i+T_0]$, and $\phi(t)=0$ otherwise. $t_i$ is the initial time of the ramping, and $T_0$ is its duration, after which the external driving accumulates an additional $2\pi$ phase thus the Hamiltonian is identical the one before the ramping.  We assume prior to the ramping ($t<t_i$), the system is in one of the DTC phase, which is ``excited'' by the additional time-dependent perturbation  provided by ramping.  After the ramping ($t>t_i+T_0$), it still takes some time for the system to relax before it enters into another  dynamical regime.

In the following, we will study both the long-time and transient dynamics of the system and demonstrate  their  dependence on the ramping rates $\frac{2\pi}{T_0}$. To this end, we fix the initial states ($V_i=10J$) and all  other parameters ($V_f=3J$, $\delta=0.8J$) except the duration of  ramping $T_0$. In the limit of $T_0=0$ the periodic driving is abruptly changed by a phase of $2\pi$ at $t=t_i$, thus the Hamiltonian is identical to that without ramping, so is the long-time dynamics. For a rapid ramping, the system exhibits a similar long-time dynamics, but with a weaker amplitude, as shown in Fig.\ref{fig:fig2} (c). In the opposite limit of slow ramping, the peak positions of $m(t)$ are pinned to those of $V(t)$. Thus, after the ramping, the phase of $V(t)$ is pushed forward by $2\pi$, so is $m(t)$. However, due to the period doubling feature of the DTC,  $m(t)$ is only shifted by a half-period, and thus falls into the other degenerate state that differs from the one before the ramping, although  $V(t)$ remains unchanged. Thus, by slowly ramping the driving phase, one can induce a dynamical transition   between two ``degenerate'' DTC phases. On the contrary, for a fast ramping, the system cannot follow $V(t)$ ``adiabatically'', thus finally relaxes to a DTC phase similar to the original one. These two distinct dynamical behaviors of  slow and fast ramping indicate a transition between them.  From Fig.\ref{fig:fig2} (d), we can find that this transition occurs suddenly at $T_0\approx2.41(1)$ and there is no crossover or intermediate regime between them.

{\it Critical behavior:} For an ideal DTC phase, the peak positions of $m(t)$ are supposed to be pinned to those of $V(t)$, whereas in realistic situations, there is always some relative displacement, which could be used to study the properties of the transition.  For each peak of $m(t)$, we define its relative displacement as $S_n(t)=|P_n(t)-2j_n|$, where $P_n(t)$ is the position of the n-th peak of m(t), $j_n$ is an integer number indicating the peak position of $V(t)$ minimizing $S_n(t)$. From Fig.\ref{fig:fig3} (a) and (b), we can find that in the DTC phase the peak positions of $m(t)$ are close to those of $V(t)$, thus one can use $S_n(t)$ to distinguish the two $Z_2$ symmetry breaking DTC phases. For solution A (B), $S_n(t)$ is close to 0 (1).  $S_n(t)$ for various $T_0$ is plotted in Fig.\ref{fig:fig3}, from which we can find that the transition between two different DTC phases can only be observed for $T_0>2.41$. It is also interesting to notice that the closer the system approaches the transition point, the longer it takes for the system to relax. It appears that the  transition duration D (or relaxation time) diverges algebraically near the critical point $D\sim |T_0^c-T_0|^{-\eta}$ with $\eta=0.96(5)$, as shown the inset in Fig.\ref{fig:fig3} (a), a reminiscence of the critical slowing down phenomena.

 \begin{figure}[htb]
\includegraphics[width=0.99\linewidth,bb=90 75 1250 550]{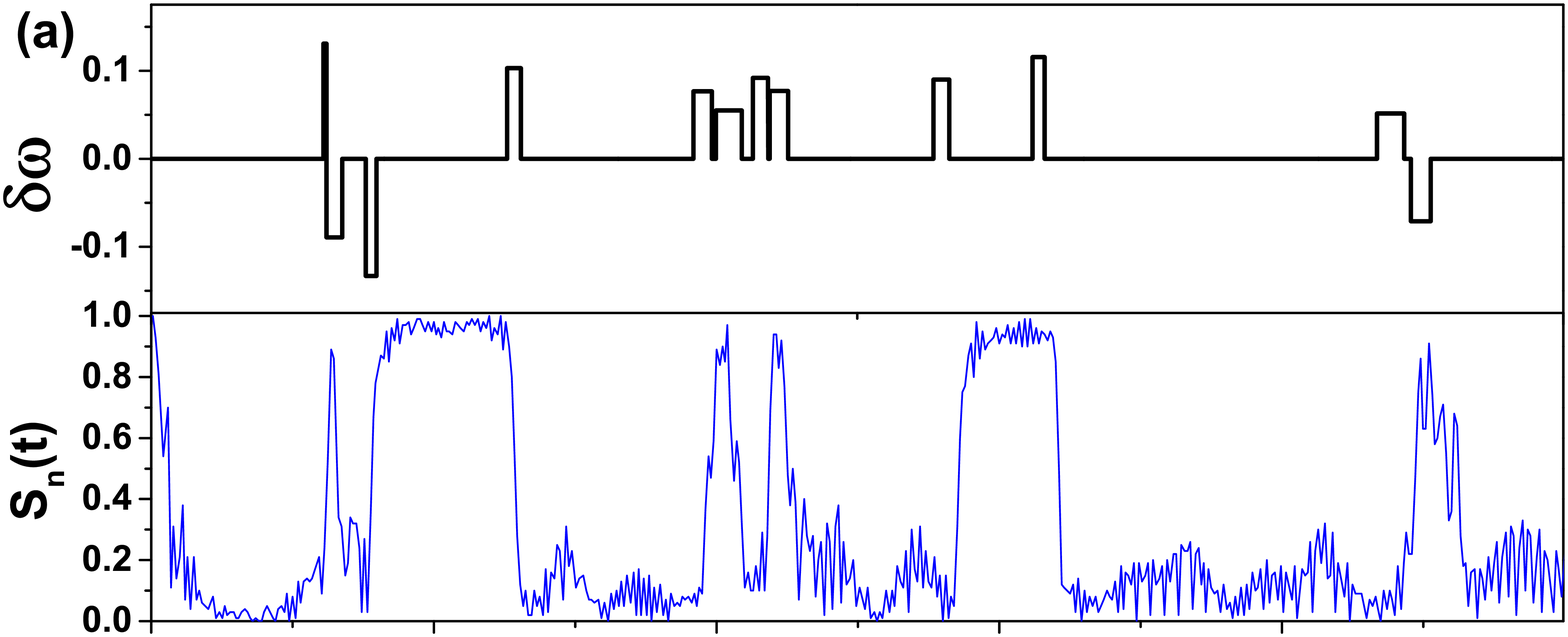}
\includegraphics[width=0.99\linewidth,bb=90 54 1250 550]{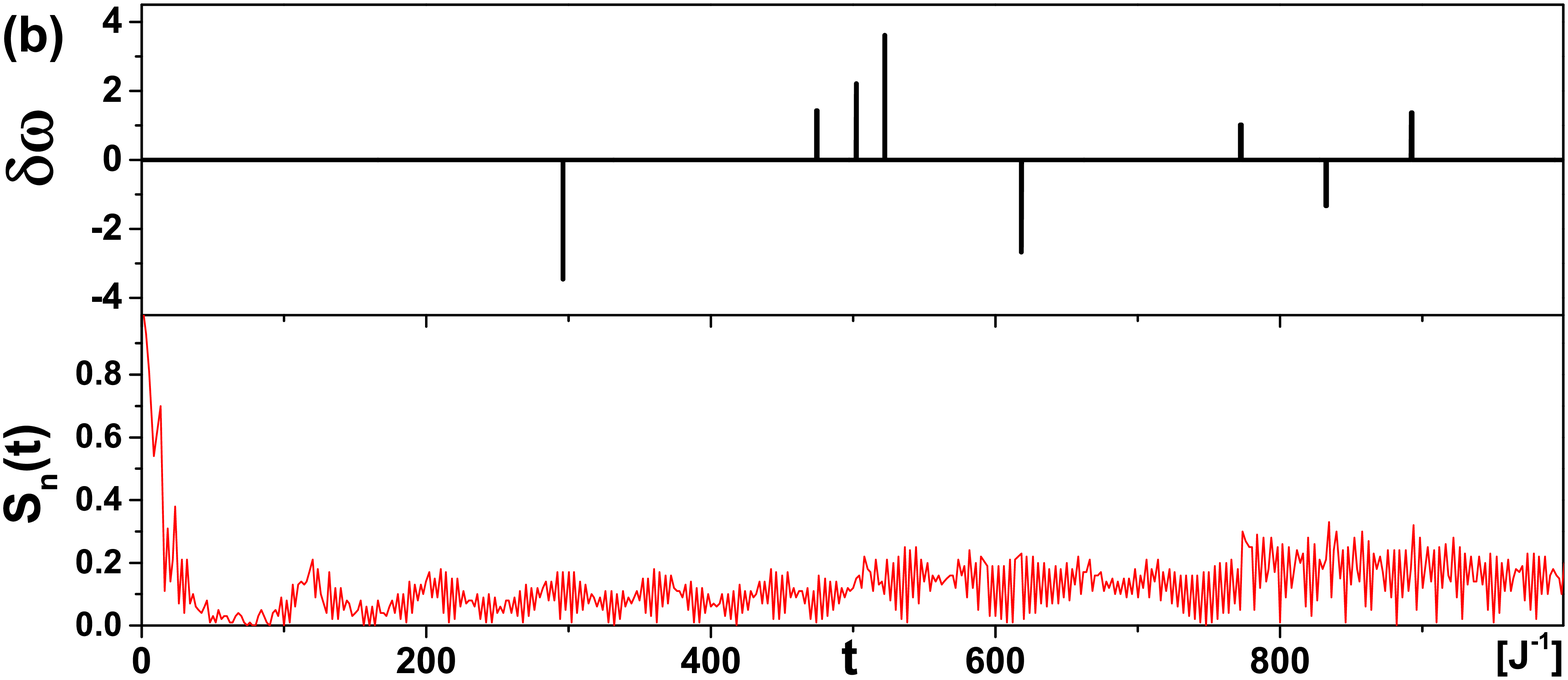}
\caption{(Color online).
Dynamics of $S_n(t)$ in the presence of two typical stochastic sequences of phase ramping  with $\alpha=0.01$,  where the driving frequency randomly fluctuates  within the regime (a) $|\delta\omega|\in [0.05,0.2]$ and (b) $|\delta\omega|\in [1,5]$. } \label{fig:fig4}
\end{figure}

{\it Stochastic sequences of phase ramping and its implication to noisy systems.} Up to now, we focus on the effect of a single phase ramping protocol on the DTC phases, and find a dynamical transition between two  DTC phases can only be induced by a sufficiently slow ramping. What's its implication to the stability of a DTC phase ({\it e.g.} its robustness against thermal fluctuation or external noise)?  To address this issue, we perform a stochastic sequence of phase ramping processes, whose initial time and duration are chosen randomly. Since the proposed phase ramping protocol in Eq.(\ref{eq:ramping}) equals to a shift of the driving frequency from $\omega=1(2\pi)$ to $\omega'=[1+\delta\omega] (2\pi)$ with $\delta\omega=\frac 1{T_0}$, a stochastic sequence of phase ramping processes with random durations ($T_0$) can be considered as a telegraph-like fluctuation of the driving frequency (see Fig.\ref{fig:fig4} for an instance), which can  be considered as a consequence of the perturbation from external noises.

In the following, we will consider the situation where the randomly-located ramping-induced-frequency fluctuations are rare events (the probability of finding such a object per unit time $\alpha\ll 1$). We also assume that for each ramping process in a stochastic sequence, its duration and the corresponding frequency fluctuation is randomly chosen within the regime $|\delta\omega|\in [\omega_1,\omega_2]$.  Two typical random sequences of frequency fluctuation and the corresponding dynamics of $S_n(t)$ are plotted  in Fig.\ref{fig:fig4}, from which we can find that a low-frequency fluctuation will induce a random sequence of dynamical transitions between DTC phases (Fig.\ref{fig:fig4} (a)). Even though for such a single stochastic sequence, the DTC order persists and its order parameter keeps oscillating between those two DTC phases, an ensemble average over all the noisy trajectories (stochastic sequences) will lead to an exponential decay of the DTC order parameter in time, which indicates the instability of the DTC phase against such noisy perturbations. On the contrary, the high-frequency fluctuation will not induce the dynamical transition, thus preserve the long-range DTC order (Fig.\ref{fig:fig4} (b)). It agrees with our result obtained from a single ramping process: there is a frequency threshold for a phase ramping protocol to induce a transition between the DTC phases.

Notice that this kind of particular telegraph frequency fluctuation  is different from those encountered in the realistic environment, thus an interesting question is whether our conclusion  can apply to DTC phases with more realistic noises, which usually  perturb the systems continuously thus can not be considered as rare events. Another important question is the effect of the thermal fluctuations induced by coupling the system to a thermal bath. Different from the noisy systems considered above, a thermal bath does not only introduce noisy fluctuations, but also dissipations. Since our original Ham.(\ref{eq:Ham}) is a quantum many-body model, its dynamics in the presence of a thermal bath  is beyond the scope of this work and will be left for the future. In general, dissipation tends to balance the noise effect, thus makes the DTC phases more stable compared to the cases with only noisy driving.

{\it Experimental realization and detection:} The proposed model can be realized in an experimental setup similar with to the one in Ref.\cite{Landig2016}.  In terms of  detections,  the CDW order parameter can  be directly measured  using the superlattice band-mapping technique\cite{Trotzky2012}, or indirectly through the heterodyne detection\cite{Landig2016,Hruby2018}. For an optical lattice with $J\approx 400$Hz, one can estimate that the typical oscillation period is in the order of a magnitude of $10$ms, which is a measurable time scale in the current experimental setup. The major obstacle for the  experimental observation is the dephasing mechanisms caused by various experimental imperfections including the dissipation (particle loss) or decoherence effect, thermal fluctuation and the extra heating effect generated by our driving protocols,  all of which will lead to the damping of oscillations. Thus, the major experimental challenge  is to control the dephasing rate and make it considerably smaller than the typical energy scale of the Hamiltonian, which enables us to observe and analyze the dynamics of  persistent oscillations before they damp.

{\it Conclusion and outlook:} In conclusion, we propose a general protocol to realize a transition between $Z_2$ symmetry breaking DTC phases, a temporal analogue  of the soliton excitation in 1D CDW system.  A dynamical critical point  has been observed, at which the transition duration diverges. Future developments will include the analysis of the effect of the time-dependent perturbation on other time crystal phases with different translational symmetry breaking. For instance, besides the DTC phase, we also observe a time crystal phase whose period is incommensurate with that of the driving\cite{Supplementary}, dubbed ``time quasicrystal''\cite{Autti2018}. By making an analogy to the incommensurate CDW states, we expect that the dynamical response of this time quasicrystal  may be significantly different from that of the DTC. Furthermore, in the absence of periodic driving, our model could manifest a persistent oscillation in its quantum quench dynamics\cite{Chen2020},  which represents a TC phase with broken symmetry with respect to continuous temporal translations. It is interesting to study the dynamical response of such continuous TC phase to time-dependent perturbation, which should  differ from excitation in the dissipative time crystal\cite{Hayata2018} since our system is disspationless.

\begin{acknowledgements}
{\it Acknowledgments}.---We  acknowledge support by the National Key Research and Development Program of China (Grant No. 2016YFA0302001), NSFC of  China (Grant No. 11674221, No.11574200),  Shanghai Municipal Science and Technology Major Project (Grant No.2019SHZDZX01) and the Program Professor of Special Appointment (Eastern Scholar) at Shanghai Institutions of Higher Learning.
\end{acknowledgements}

\newpage

\begin{center}
\textbf{\Large{Supplemental material}}
\end{center}

\section{Validity of the time-dependent self-consistent mean-field method}
It is well known that for a model with infinitely long-range  interaction, the mean-field method provides an exact treatment for the equilibrium properties in the thermodynamic limit. In the following, we will prove that this is also the case for the dynamical problems.

We denote $\mathfrak{F}$ is the Fock basis of hard-core bosons (e.g. $|0110\cdots 11\rangle$), and divide the evolution period $[0,t]$ into M slices with $\Delta t=t/M$. In the limit of $M\rightarrow \infty$, the path integral expression of the  propagators can be written as:
\begin{equation}
\mathcal{K}=\sum_{\{\mathfrak{F}^1,\cdots\mathfrak{F}^{M-1}\}}\langle \mathfrak{F}^0|e^{-i\Delta t H(t_1)}|\mathfrak{F}^1\rangle\langle \mathfrak{F}^1|\cdots e^{-i\Delta t H(t_M)}|\mathfrak{F}^M\rangle \label{eq:path}
\end{equation}
where $|\mathfrak{F}^j\rangle$ is a Fock basis at the $j$-th time slice ($j=0,\cdots M$), with $\sum_{\{\mathfrak{F}^j\}}|\mathfrak{F}^j\rangle\langle \mathfrak{F}^j|$ a unit matrix. We further divide our Hamiltonian as $H=\hat{T}+\hat{V}$ where $\hat{T}=-J\sum_i (\hat{b}_i^\dag \hat{b}_{i+1}+h.c)$, and the interacting energy can be rewritten as:
\begin{equation}
\hat{V}=LV(t)[\frac 1L\sum_i(-1)^i\hat{n}_i]^2.
\end{equation}
By performing Suzuki-Trotter decomposition, one can obtain that
\begin{equation}
e^{-i\Delta tH}=e^{-i\Delta t \hat{T}}e^{-i\Delta t\hat{V}}+\mathcal{O}(\Delta t^2)
\end{equation}
Therefore, in the limit of $\Delta t\rightarrow 0$, the propagator in Eq.(\ref{eq:path}) can be expressed as:
\begin{equation}
\mathcal{K}=\sum_{\{\mathfrak{F}^1,\cdots\mathfrak{F}^{M-1}\}} \prod_{j=1}^{M-1}T_{j,j+1}e^{-i\Delta t V_{\{\mathfrak{F}^j\}}(t_j)}  \label{eq:path2}
\end{equation}
where $T_{j,j+1}=\langle \mathfrak{F}^j|e^{-i\Delta t \hat{T}}|\mathfrak{F}^{j+1}\rangle$, and  $V_{\{\mathfrak{F}^j\}}(t_j)=\langle \mathfrak{F}^j|e^{-i\Delta t \hat{V}}|\mathfrak{F}^{j}\rangle=e^{-i\Delta t LV(t_i)\mathcal{S}\{\mathfrak{F}^j\}^2}$, where $\mathcal{S}\{\mathfrak{F}^j\}=\frac 1L\sum_i(-1)^in_i|_{\{n_i\}\in\{\mathfrak{F}^j\}}$.

The quadratic part in Eq.(\ref{eq:path2}) can now be decoupled using the Hubbard-Stratonovic transformation by introducing auxillary fields $\{m^j\}$ with $j=1,M-1$ as:
\begin{equation}
\mathcal{K}= \prod_{j=1}^{M-1}\sum_{\{\mathfrak{F}^j\}}\int \frac{dm^j}{\mathcal{N}} T_{j,j+1}e^{-i\Delta t LV(t_j)[(m^j)^2-2m^j\mathcal{S}\{\mathfrak{F}^j\}]}  \label{eq:path3}
\end{equation}
where $\mathcal{N}$ is a normalization factor to keep the identity of the Gaussian integral. In the thermodynamic limit $L\rightarrow \infty$, Eq.(\ref{eq:path3}) indicates that the saddle point approximation $\frac{\delta \mathcal{L}[m^j]}{\delta m^j}=0$ with  $\mathcal{L}[m^j]=V(t_j)[(m^j)^2-2m^j\mathcal{S}\{\mathfrak{F}^j\}]$, becomes exact,  which  means that the auxillary fields  $\{m^j\}$ are given by the CDW order parameter of the state as $m^j=\mathcal{S}\{\mathfrak{F}^j\}$. As a consequence, For the time evolution of a state $|\psi\rangle$   in each infinitesimal time step ($\Delta t\rightarrow 0$),  the evolution operator $e^{iH(t)\delta t}$  are equal to the one after mean-field approximation  $e^{i\bar{H}(t,m(t))\Delta t}$   with a time-dependent CDW order parameters $m(t)$, which is determined self-consistently during the time evolution by the saddle point method.

 \begin{figure}[htb]
\includegraphics[width=0.99\linewidth,bb=84 33 806 429]{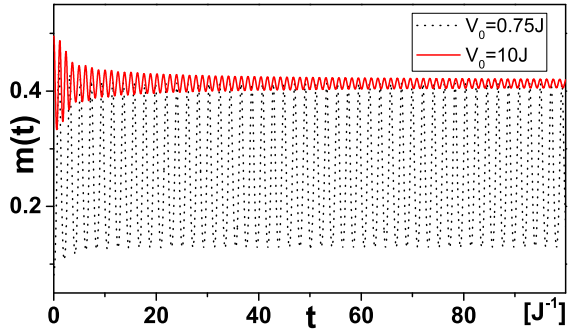}
\caption{(Color online).
Quench dynamics of $M(t)$ calculated by the self-consistent mean-field method with different $V_i$ and the parameters $V_f=3J$, $L=5000$, $\Delta t=10^{-5}$, $\delta=0$;} \label{fig:SM1}
\end{figure}
\section{Quantum quench dynamics without driving}

In this section, we study quantum quench dynamics of the model ($\delta=0$) by choosing an initial state as the ground state of $H$ with $V(t)=V_i$, and suddenly change the interaction strength to a different value $V_f$ and let the system evolve under this new Hamiltonian. The long-time dynamics of this model has been analyzed in Ref.[1], here we will outline the mean-field results of Ref.[1]. We plot the time evolution of the CDW order parameter $m(t)=\frac 1L\sum_i (-1)^i \langle n_i\rangle$ starting from two different initial states  in Fig.\ref{fig:SM1}, from which we can find  two distinct dynamical behaviors  even though the evolution Hamiltonian are the same ($V_f=3J$). For the quench dynamics from the initial state with $V_i=10J$, $m(t)$ will converge to a constant after sufficiently long time, while for the other case with  $V_i=0.75J$, it persistently oscillates with a period that spontaneously emerges during the quench dynamics.

 \begin{figure}[htb]
\includegraphics[width=0.99\linewidth,bb=70 150 765 999]{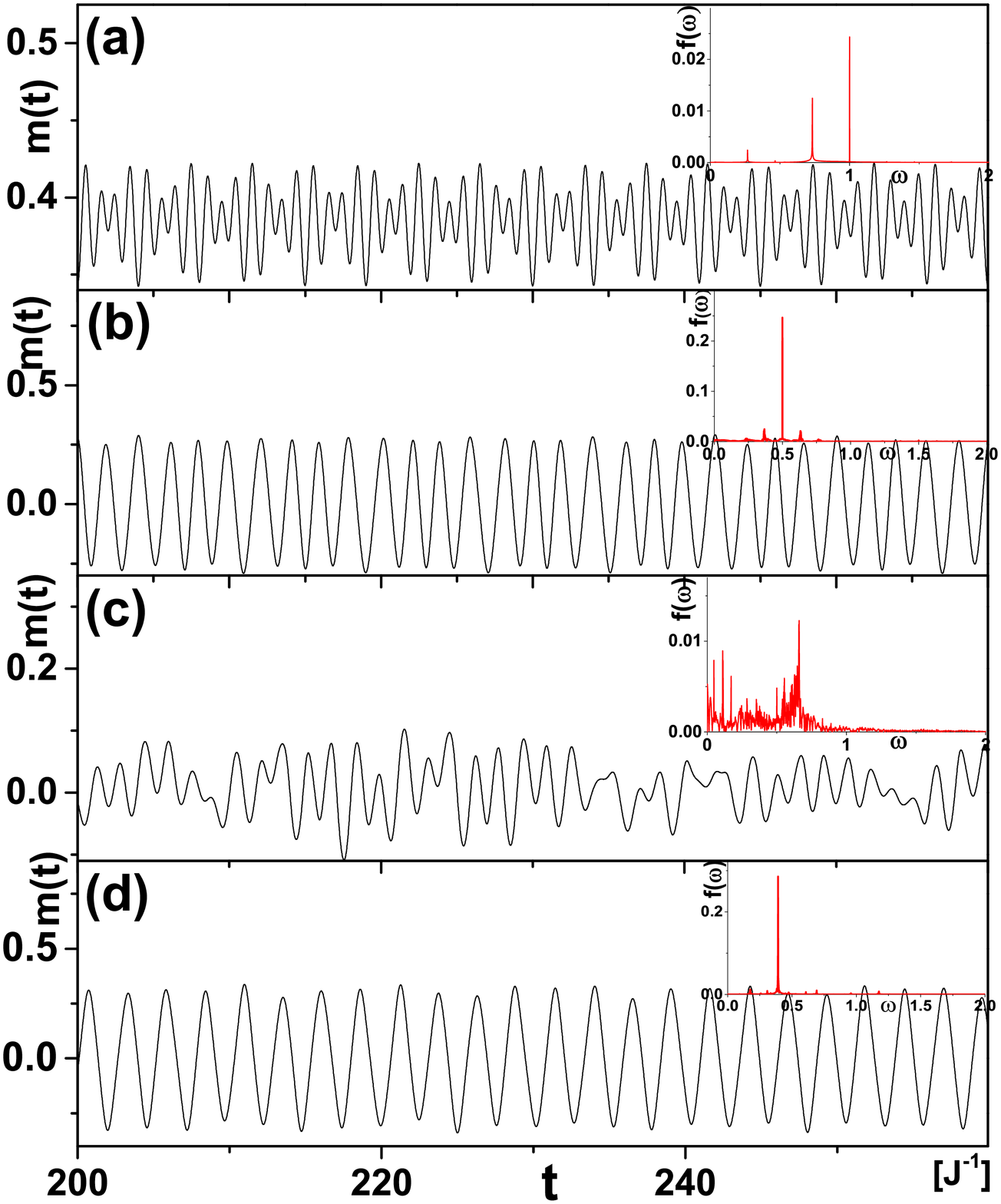}
\caption{(Color online).
Dependence of periodically driven dynamics of $M(t)$ on driving amplitudes [(a) $\delta=0.3J$; (b) $\delta=0.8J$; (c) $\delta=8J$;] and  initial states [(d) $V_i=0.75J$, $\delta=0.8J$]. Other parameters are chosen as $V_i=10J$ for (a)-(c), $L=5000$ and $\Delta t=10^{-5}J^{-1}$ for (a)-(d). } \label{fig:SM2}
\end{figure}

 \begin{figure}[htb]
\includegraphics[width=0.99\linewidth,bb=90 18 765 369]{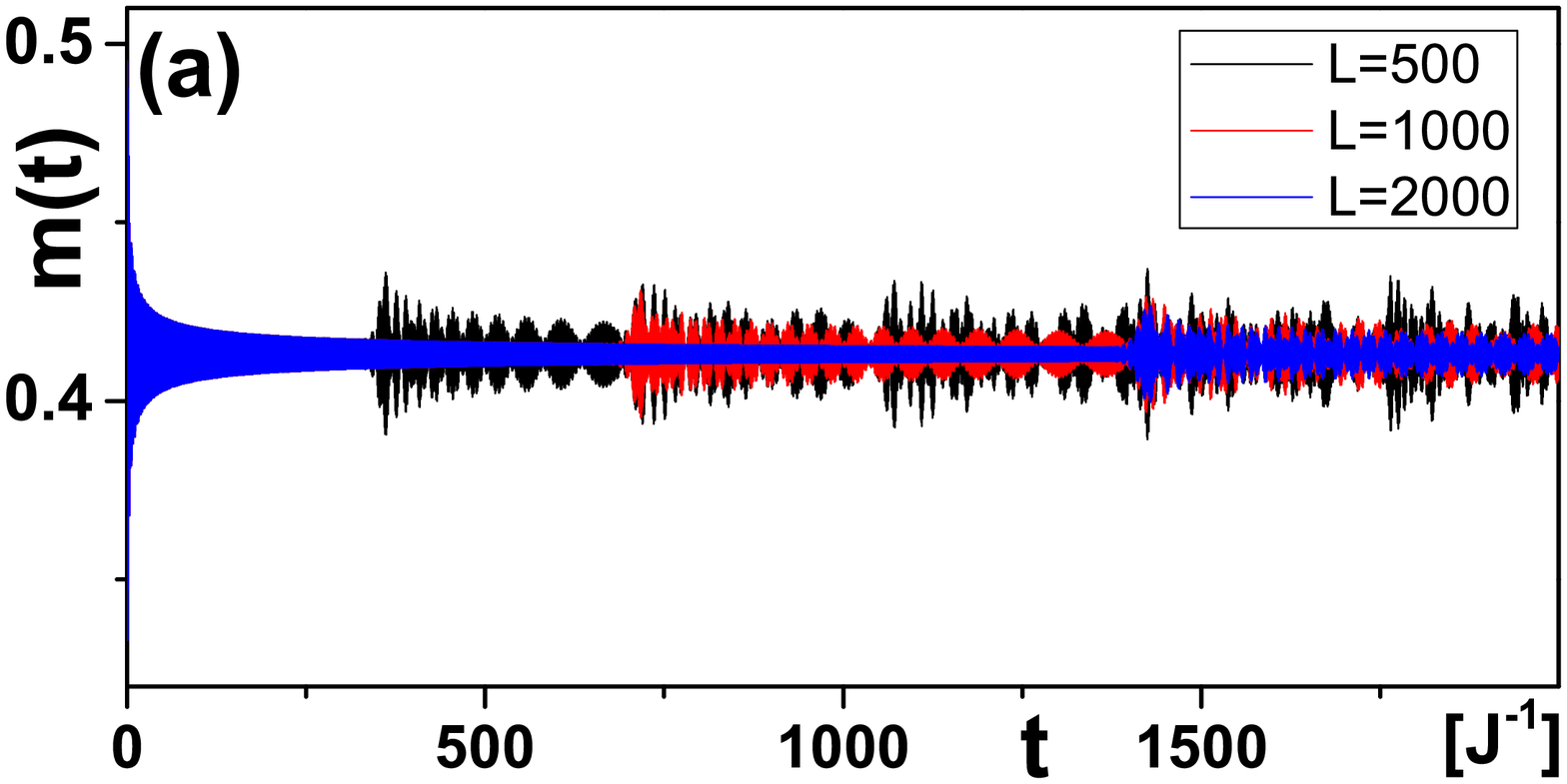}
\includegraphics[width=0.99\linewidth,bb=90 18 765 369]{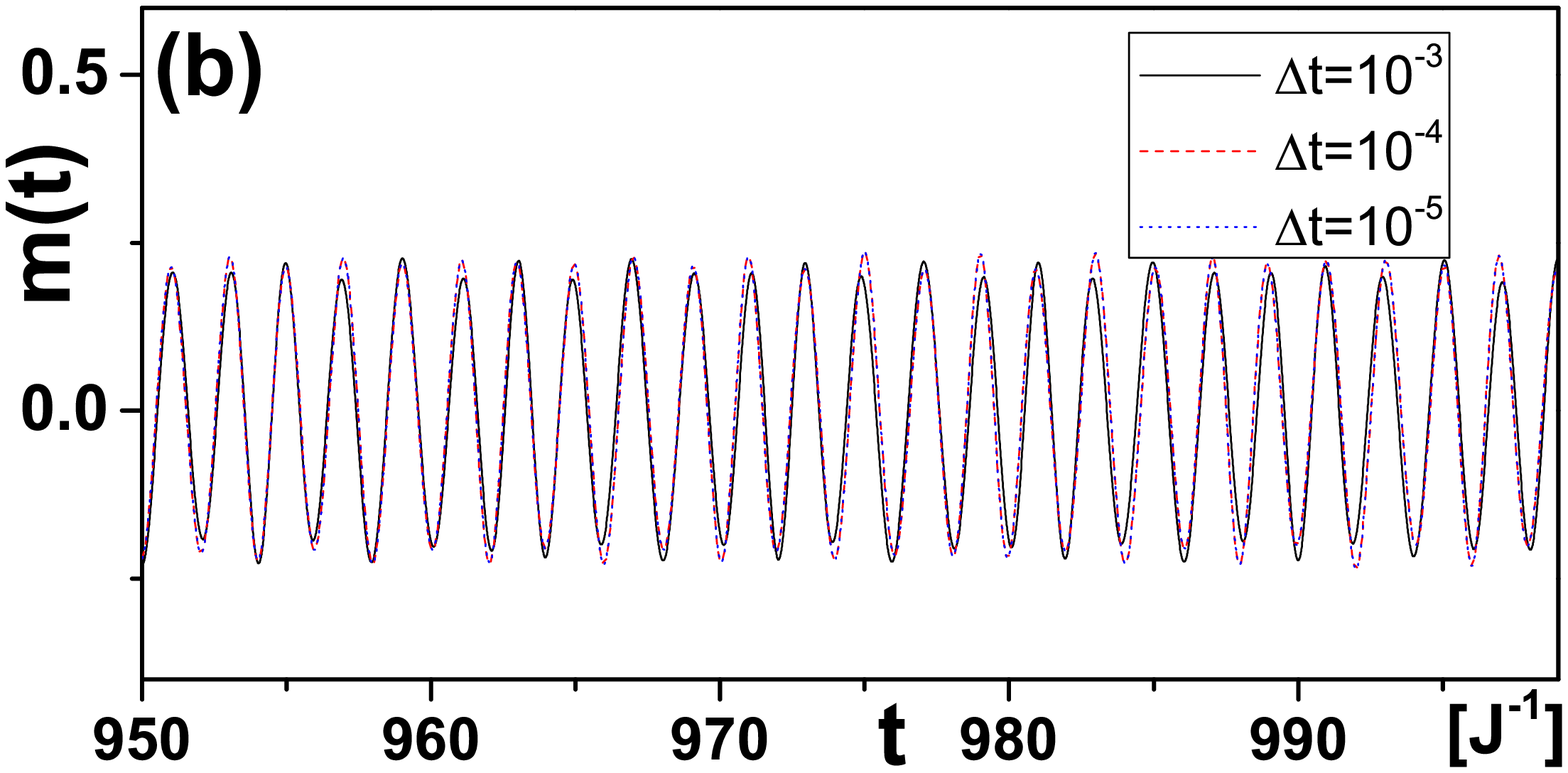}
\caption{(Color online).
(a) Quench dynamics of $M(t)$ with different system size $L$ and the parameters $\delta=0$, $\Delta t=10^{-5}$; (b) Periodically driven dynamics with different $\Delta t$ and $\delta=0.8J$, $L=5000$; $V_i=10J$ and $V_f=3J$ for (a) and (b)} \label{fig:SM3}
\end{figure}

\section{Periodically driven dynamics}
The periodically driven dynamical behavior in our model is very rich, while in the main text we only focus on the time crystal phase. Here we list several typical long-time behaviors, even though it is difficult to enumerate all the possibilities.

In Fig.(\ref{fig:SM3}), we plot the long-time dynamics of m(t) with  different   driving amplitude $\delta$ and initial states $V_i$. For Fig.\ref{fig:SM3} (a)-(c), we fixed the initial state as the ground state of the Hamiltonian with $V_i=10J$, while for Fig.\ref{fig:SM3} (d), we change it to $V_i=0.75J$.  As shown in Fig.\ref{fig:SM3} (a), for a weak periodical driving, the long-time dynamics becomes a persistent quasi-periodic oscillation with a multi-period structure, which can be reflected in its Fourier spectrum $f(\omega)=\frac 1{T}\int dt e^{i\omega t} m(t)$. As shown in the inset of Fig.\ref{fig:SM3} (a), the Fourier spectrum $f(\omega)$ exhibits sharp peaks at different characteristic frequency, while the dominant one locates at the place exactly the same with the external driving frequency ($\omega=1$), which indicates $m(t)$ is synchronous with the external periodic driving. For an intermediate driving amplitude, for instance $\delta=0.8J$ as shown in Fig.\ref{fig:SM3} (b), the peak position of $f(\omega)$ has been shifted from $1$ to $0.5$, which indicates a period-doubling time crystal phase as we studied in the main text.   When we further increase $\delta$ to a strongly driven regime ({\it e.g.} $\delta=8J$ as shown in Fig.\ref{fig:SM3} c), the frequency spectrum exhibits a broad distribution, a signature of a chaotic dynamics without a dominant time scale. If we start from an different initial state, for instance, the ground state of $V_i=0.75J$ as shown in Fig.\ref{fig:SM3} (d), we can find $m(t)$ exhibits an persistent oscillation with a frequency $\omega=0.46$, which has nothing to do with the external driving frequency.

\section{Finite size effect and Convergence of the numerical results}

In our numerical simulations, we choose the lattice size as $L=5000$. Here we will show that within our simulation period ($0<t<2000J^{-1}$), such a system size is sufficiently large that the finite-size effect can be neglected. To study the finite-size effect, we calculate $m(t)$ for smaller systems. Take the quantum quench dynamics for an instance ($V_i=10J$, $V_j=3J$, $\delta=0$), where $m(t)$ is supposed to decay to a constant value after sufficiently long time. However,  as shown in Fig.\ref{fig:SM3} (a), for a small system, a revival of oscillation will occur at the time $t_r\sim \frac{L}{\sqrt{2}J}$, which is the consequence of the finite size effect. Therefore, as long as our simulation time $(2000J^{-1}$) is smaller than the revival time $5000/(\sqrt{2}J)$, the finite size effect in our simulation can be neglected.

Another parameter in our simulation is the time step $\Delta t$, which is chosen to be $\Delta t=10^{-5}J^{-1}$ in our simulation. In  Fig.\ref{fig:SM3} (b), we check the results with different $\Delta t$, which indicates that our results are sufficiently converged for $\Delta t=10^{-5}J^{-1}$.

 \begin{figure}[htb]
\includegraphics[width=0.95\linewidth,bb=147 54 950 528]{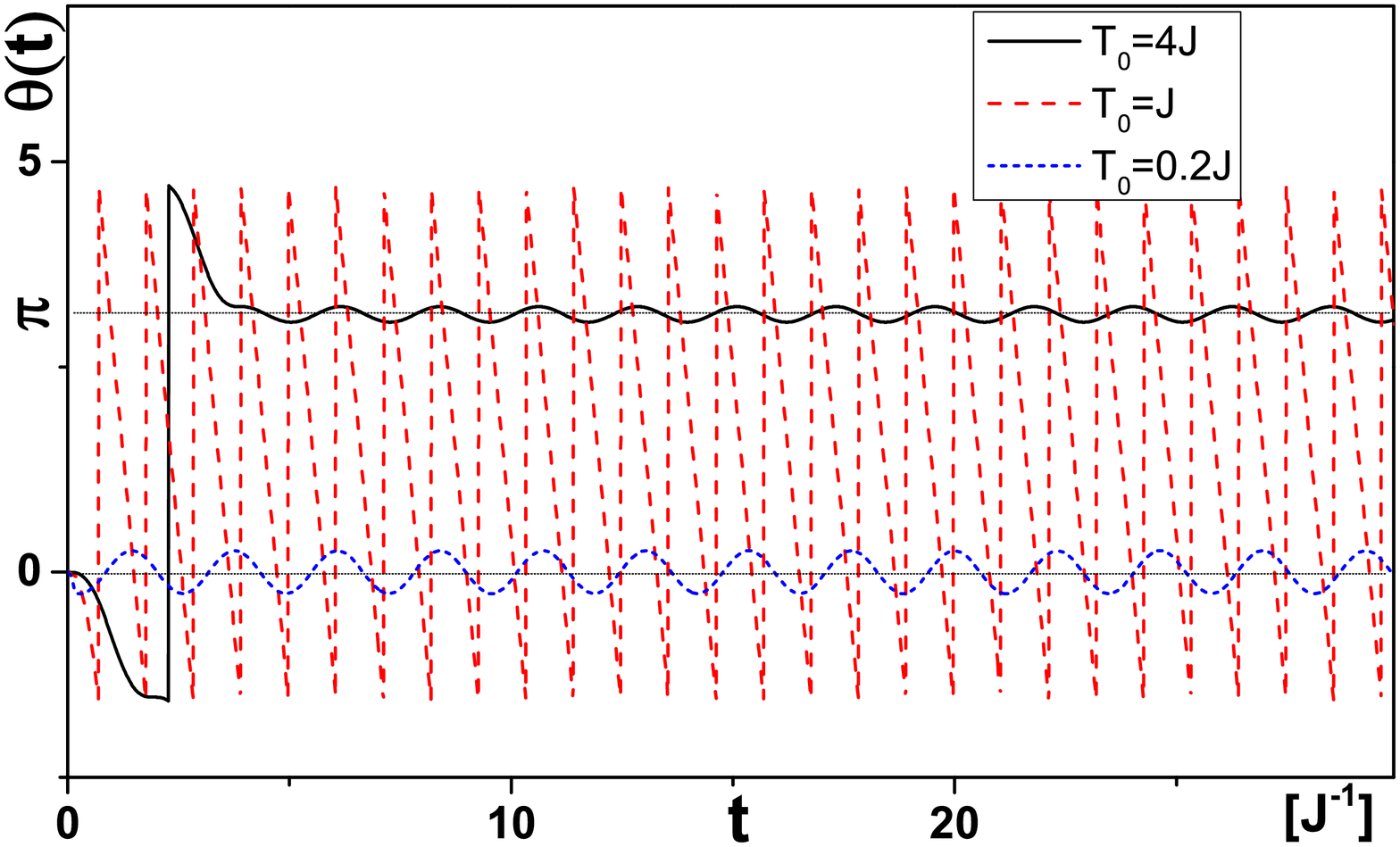}
\caption{(Color online). Dynamics of $\tilde{\theta}(t)$ predicted by the Ham.(\ref{eq:Hamt2}) in the presence of fast, intermediate and slow  ramping with parameters $\alpha=\gamma=J$, and $\tilde{\theta}\in [-\frac\pi 2,\frac{3\pi}2]$ with PBC.} \label{fig:fig4}
\end{figure}

\section {Dynamical transitions in other DTC models}
The proposed driving protocol is general in the sense that it can induce similar dynamical transitions between the DTC phases in other models. As an example, we consider a classical nonlinear oscillator with the Hamiltonian:
\begin{equation}
H(t)=\alpha I-\gamma I^2+  \tilde{V}(t) I\cos^2\theta \label{eq:Hamt}
\end{equation}
where we assume $p=\sqrt{I}\sin\theta$, $q=\sqrt{I}\cos\theta$ with p/q the momentum/coordinate of the oscillator. The nonlinearity of the oscillator is encoded in $I^2$.  $\tilde{V}(t)=\cos[2\pi (t+\phi(t))]$ with $\phi(t)$ obeys the same phase ramping protocol  proposed above. In the absence of ramping ($\phi(t)=0$), it is known that such a driven oscillator could support period doubling solution $q(t)\sim \cos\pi t$ in certain parameter regime\cite{Yao2020}.  By imposing phase ramping on top of the period doubling dynamics, we assume that $q(t)\sim \cos[\pi t+\tilde{\theta}(t)]$, where $\tilde{\theta}(t)$ represents the phase deviation from the period doubling solution, which resembles the relative displacement $S_n$ defined above. By substituting the $q(t)$ into Ham.(\ref{eq:Hamt}) and omitting the fastly oscillating terms (e.g. $\cos(4\pi t+2\pi\phi(t))$, one can obtain an effective Hamiltonian in terms of $\tilde{\theta}(t)$ as:
\begin{equation}
\tilde{H}(t)=\alpha I-\gamma I^2 +I\cos[2\tilde{\theta}-2\pi\phi(t)] \label{eq:Hamt2}
\end{equation}

We first focus on the case without ramping, where Ham.(\ref{eq:Hamt2}) can be considered as a time independent effective Floquet description of Ham.(\ref{eq:Hamt}) to the 1st order of Magnus expansion\cite{Yao2020}. The corresponding energy landscape in $(\tilde{\theta},I)$ plane is an inverted double well potential  with two degenerate maxima at $I=\frac{\alpha}{2\gamma}$, and $\tilde{\theta}=0$ or $\pi$, which resembles  the two DTC solutions discussed above ($S_n=0$ or 1). Assuming initially the system locates at one of the maxima ($I(t=0)=\frac{\alpha}{2\gamma}$, $\tilde{\theta}(t=0)=0$), we expect that small perturbations will lead to bounded orbits   around such maximum in the $(\tilde{\theta},I)$ plane, while large perturbation may induce a tunneling between the two maxima. This expectation agrees with our numerical results based on Ham.(\ref{eq:Hamt2}). As shown in Fig.\ref{fig:fig4}, for a quick ramping, $\tilde{\theta}$ has no time to follow the movement of the potential, and thus it will maintain in its original maximum. However, a sufficiently slow shift of the potential will drive $\tilde{\theta}$ from $0$ to $\pi$. For an intermediate ramping rate, $\tilde{\theta(t)}$ keeps oscillating between the two maxima.

Even though the dynamical behaviors of such a simple nonlinear model resemble those observed in our many-body Hamiltonian, there is an important discrepancy: in the many-body Hamiltonian (Ham.(1) in the main text), we find a finite regime of the ramping velocity where the system can oscillate between these two maxima without decay, while in the many-body system,  such a finite regime shrinks into a  point: the system will finally relax to one of the DTC phase except in one critical point. Furthermore, the results of such a nonlinear oscillator indicate that the phase ramping protocol proposed above may be general in the sence that it can also induce the transition between the $Z_2$ symmetry breaking DTC phases in other TC models  (the Floquet TC for instance).  However, whether the ``critical'' behavior (if exists)  is universal or model-dependent is an  interesting open question that deserves further numerical or experimental studies.


\begin{thebibliography}{42}
\expandafter\ifx\csname natexlab\endcsname\relax\def\natexlab#1{#1}\fi
\expandafter\ifx\csname bibnamefont\endcsname\relax
  \def\bibnamefont#1{#1}\fi
\expandafter\ifx\csname bibfnamefont\endcsname\relax
  \def\bibfnamefont#1{#1}\fi
\expandafter\ifx\csname citenamefont\endcsname\relax
  \def\citenamefont#1{#1}\fi
\expandafter\ifx\csname url\endcsname\relax
  \def\url#1{\texttt{#1}}\fi
\expandafter\ifx\csname urlprefix\endcsname\relax\def\urlprefix{URL }\fi
\providecommand{\bibinfo}[2]{#2}
\providecommand{\eprint}[2][]{\url{#2}}

\bibitem[{\citenamefont{Anderson}(1997)}]{Anderson1997}
\bibinfo{author}{\bibfnamefont{P.}~\bibnamefont{Anderson}},
  \emph{\bibinfo{title}{Concepts in Solids}} (\bibinfo{publisher}{~World
  Scientific}, \bibinfo{year}{1997}).

\bibitem[{\citenamefont{Eisert et~al.}(2015)\citenamefont{Eisert, Friesdorf,
  and Gogolin}}]{Eisert2015}
\bibinfo{author}{\bibfnamefont{J.}~\bibnamefont{Eisert}},
  \bibinfo{author}{\bibfnamefont{M.}~\bibnamefont{Friesdorf}},
  \bibnamefont{and} \bibinfo{author}{\bibfnamefont{C.}~\bibnamefont{Gogolin}},
  \bibinfo{journal}{Nature Phys} \textbf{\bibinfo{volume}{11}},
  \bibinfo{pages}{124} (\bibinfo{year}{2015}).

\bibitem[{\citenamefont{Mankowsky et~al.}(2014)\citenamefont{Mankowsky, Subedi,
  Forst, Mariager, Chollet, Lemke, Robinson, Glownia, Minitti, Frano
  et~al.}}]{Mankowsky2014}
\bibinfo{author}{\bibfnamefont{R.}~\bibnamefont{Mankowsky}},
  \bibinfo{author}{\bibfnamefont{A.}~\bibnamefont{Subedi}},
  \bibinfo{author}{\bibfnamefont{M.}~\bibnamefont{Forst}},
  \bibinfo{author}{\bibfnamefont{S.~O.} \bibnamefont{Mariager}},
  \bibinfo{author}{\bibfnamefont{M.}~\bibnamefont{Chollet}},
  \bibinfo{author}{\bibfnamefont{H.~T.} \bibnamefont{Lemke}},
  \bibinfo{author}{\bibfnamefont{J.~S.} \bibnamefont{Robinson}},
  \bibinfo{author}{\bibfnamefont{J.~M.} \bibnamefont{Glownia}},
  \bibinfo{author}{\bibfnamefont{M.~P.} \bibnamefont{Minitti}},
  \bibinfo{author}{\bibfnamefont{A.}~\bibnamefont{Frano}},
  \bibnamefont{et~al.}, \bibinfo{journal}{Nature}
  \textbf{\bibinfo{volume}{516}}, \bibinfo{pages}{71} (\bibinfo{year}{2014}).

\bibitem[{\citenamefont{Wilczek}(2012)}]{Wilczek2012}
\bibinfo{author}{\bibfnamefont{F.}~\bibnamefont{Wilczek}},
  \bibinfo{journal}{Phys. Rev. Lett.} \textbf{\bibinfo{volume}{109}},
  \bibinfo{pages}{160401} (\bibinfo{year}{2012}).

\bibitem[{\citenamefont{Shapere and Wilczek}(2012)}]{Shapere2012}
\bibinfo{author}{\bibfnamefont{A.}~\bibnamefont{Shapere}} \bibnamefont{and}
  \bibinfo{author}{\bibfnamefont{F.}~\bibnamefont{Wilczek}},
  \bibinfo{journal}{Phys. Rev. Lett.} \textbf{\bibinfo{volume}{109}},
  \bibinfo{pages}{160402} (\bibinfo{year}{2012}).

\bibitem[{\citenamefont{Li et~al.}(2012)\citenamefont{Li, Gong, Yin, Quan, Yin,
  Zhang, Duan, and Zhang}}]{Li2012}
\bibinfo{author}{\bibfnamefont{T.}~\bibnamefont{Li}},
  \bibinfo{author}{\bibfnamefont{Z.-X.} \bibnamefont{Gong}},
  \bibinfo{author}{\bibfnamefont{Z.-Q.} \bibnamefont{Yin}},
  \bibinfo{author}{\bibfnamefont{H.~T.} \bibnamefont{Quan}},
  \bibinfo{author}{\bibfnamefont{X.}~\bibnamefont{Yin}},
  \bibinfo{author}{\bibfnamefont{P.}~\bibnamefont{Zhang}},
  \bibinfo{author}{\bibfnamefont{L.-M.} \bibnamefont{Duan}}, \bibnamefont{and}
  \bibinfo{author}{\bibfnamefont{X.}~\bibnamefont{Zhang}},
  \bibinfo{journal}{Phys. Rev. Lett.} \textbf{\bibinfo{volume}{109}},
  \bibinfo{pages}{163001} (\bibinfo{year}{2012}).

\bibitem[{\citenamefont{Wilczek}(2013)}]{Wilczek2013}
\bibinfo{author}{\bibfnamefont{F.}~\bibnamefont{Wilczek}},
  \bibinfo{journal}{Phys. Rev. Lett.} \textbf{\bibinfo{volume}{111}},
  \bibinfo{pages}{250402} (\bibinfo{year}{2013}).

\bibitem[{\citenamefont{Sacha}(2015)}]{Sacha2015}
\bibinfo{author}{\bibfnamefont{K.}~\bibnamefont{Sacha}},
  \bibinfo{journal}{Phys. Rev. A} \textbf{\bibinfo{volume}{91}},
  \bibinfo{pages}{033617} (\bibinfo{year}{2015}).

\bibitem[{\citenamefont{Else et~al.}(2016)\citenamefont{Else, Bauer, and
  Nayak}}]{Else2016}
\bibinfo{author}{\bibfnamefont{D.~V.} \bibnamefont{Else}},
  \bibinfo{author}{\bibfnamefont{B.}~\bibnamefont{Bauer}}, \bibnamefont{and}
  \bibinfo{author}{\bibfnamefont{C.}~\bibnamefont{Nayak}},
  \bibinfo{journal}{Phys. Rev. Lett.} \textbf{\bibinfo{volume}{117}},
  \bibinfo{pages}{090402} (\bibinfo{year}{2016}).

\bibitem[{\citenamefont{Khemani et~al.}(2016)\citenamefont{Khemani, Lazarides,
  Moessner, and Sondhi}}]{Khemani2016}
\bibinfo{author}{\bibfnamefont{V.}~\bibnamefont{Khemani}},
  \bibinfo{author}{\bibfnamefont{A.}~\bibnamefont{Lazarides}},
  \bibinfo{author}{\bibfnamefont{R.}~\bibnamefont{Moessner}}, \bibnamefont{and}
  \bibinfo{author}{\bibfnamefont{S.~L.} \bibnamefont{Sondhi}},
  \bibinfo{journal}{Phys. Rev. Lett.} \textbf{\bibinfo{volume}{116}},
  \bibinfo{pages}{250401} (\bibinfo{year}{2016}).

\bibitem[{\citenamefont{Yao et~al.}(2017)\citenamefont{Yao, Potter, Potirniche,
  and Vishwanath}}]{Yao2017}
\bibinfo{author}{\bibfnamefont{N.~Y.} \bibnamefont{Yao}},
  \bibinfo{author}{\bibfnamefont{A.~C.} \bibnamefont{Potter}},
  \bibinfo{author}{\bibfnamefont{I.-D.} \bibnamefont{Potirniche}},
  \bibnamefont{and}
  \bibinfo{author}{\bibfnamefont{A.}~\bibnamefont{Vishwanath}},
  \bibinfo{journal}{Phys. Rev. Lett.} \textbf{\bibinfo{volume}{118}},
  \bibinfo{pages}{030401} (\bibinfo{year}{2017}).

\bibitem[{\citenamefont{Russomanno et~al.}(2017)\citenamefont{Russomanno,
  Iemini, Dalmonte, and Fazio}}]{Russomanno2017}
\bibinfo{author}{\bibfnamefont{A.}~\bibnamefont{Russomanno}},
  \bibinfo{author}{\bibfnamefont{F.}~\bibnamefont{Iemini}},
  \bibinfo{author}{\bibfnamefont{M.}~\bibnamefont{Dalmonte}}, \bibnamefont{and}
  \bibinfo{author}{\bibfnamefont{R.}~\bibnamefont{Fazio}},
  \bibinfo{journal}{Phys. Rev. B} \textbf{\bibinfo{volume}{95}},
  \bibinfo{pages}{214307} (\bibinfo{year}{2017}).

\bibitem[{\citenamefont{Gong et~al.}(2018)\citenamefont{Gong, Hamazaki, and
  Ueda}}]{Gong2018}
\bibinfo{author}{\bibfnamefont{Z.}~\bibnamefont{Gong}},
  \bibinfo{author}{\bibfnamefont{R.}~\bibnamefont{Hamazaki}}, \bibnamefont{and}
  \bibinfo{author}{\bibfnamefont{M.}~\bibnamefont{Ueda}},
  \bibinfo{journal}{Phys. Rev. Lett.} \textbf{\bibinfo{volume}{120}},
  \bibinfo{pages}{040404} (\bibinfo{year}{2018}).

\bibitem[{\citenamefont{Huang et~al.}(2018)\citenamefont{Huang, Wu, and
  Liu}}]{Huang2018}
\bibinfo{author}{\bibfnamefont{B.}~\bibnamefont{Huang}},
  \bibinfo{author}{\bibfnamefont{Y.-H.} \bibnamefont{Wu}}, \bibnamefont{and}
  \bibinfo{author}{\bibfnamefont{W.~V.} \bibnamefont{Liu}},
  \bibinfo{journal}{Phys. Rev. Lett.} \textbf{\bibinfo{volume}{120}},
  \bibinfo{pages}{110603} (\bibinfo{year}{2018}).

\bibitem[{\citenamefont{Iemini et~al.}(2018)\citenamefont{Iemini, Russomanno,
  Keeling, Schir\`o, Dalmonte, and Fazio}}]{Iemini2018}
\bibinfo{author}{\bibfnamefont{F.}~\bibnamefont{Iemini}},
  \bibinfo{author}{\bibfnamefont{A.}~\bibnamefont{Russomanno}},
  \bibinfo{author}{\bibfnamefont{J.}~\bibnamefont{Keeling}},
  \bibinfo{author}{\bibfnamefont{M.}~\bibnamefont{Schir\`o}},
  \bibinfo{author}{\bibfnamefont{M.}~\bibnamefont{Dalmonte}}, \bibnamefont{and}
  \bibinfo{author}{\bibfnamefont{R.}~\bibnamefont{Fazio}},
  \bibinfo{journal}{Phys. Rev. Lett.} \textbf{\bibinfo{volume}{121}},
  \bibinfo{pages}{035301} (\bibinfo{year}{2018}).

\bibitem[{\citenamefont{Das et~al.}(2018)\citenamefont{Das, Pan, Ghosh, and
  Pal}}]{Das2018}
\bibinfo{author}{\bibfnamefont{P.}~\bibnamefont{Das}},
  \bibinfo{author}{\bibfnamefont{S.}~\bibnamefont{Pan}},
  \bibinfo{author}{\bibfnamefont{S.}~\bibnamefont{Ghosh}}, \bibnamefont{and}
  \bibinfo{author}{\bibfnamefont{P.}~\bibnamefont{Pal}},
  \bibinfo{journal}{Phys. Rev. D} \textbf{\bibinfo{volume}{98}},
  \bibinfo{pages}{024004} (\bibinfo{year}{2018}).

\bibitem[{\citenamefont{Zhu et~al.}(2019)\citenamefont{Zhu, Marino, Yao, Lukin,
  and Demler}}]{Zhu2019}
\bibinfo{author}{\bibfnamefont{B.}~\bibnamefont{Zhu}},
  \bibinfo{author}{\bibfnamefont{J.}~\bibnamefont{Marino}},
  \bibinfo{author}{\bibfnamefont{N.~Y.} \bibnamefont{Yao}},
  \bibinfo{author}{\bibfnamefont{M.~D.} \bibnamefont{Lukin}}, \bibnamefont{and}
  \bibinfo{author}{\bibfnamefont{E.~A.} \bibnamefont{Demler}},
  \bibinfo{journal}{New Journal of Physics} \textbf{\bibinfo{volume}{21}},
  \bibinfo{pages}{073028} (\bibinfo{year}{2019}).

\bibitem[{\citenamefont{Kozin and Kyriienko}(2019)}]{Kozin2019}
\bibinfo{author}{\bibfnamefont{V.~K.} \bibnamefont{Kozin}} \bibnamefont{and}
  \bibinfo{author}{\bibfnamefont{O.}~\bibnamefont{Kyriienko}},
  \bibinfo{journal}{Phys. Rev. Lett.} \textbf{\bibinfo{volume}{123}},
  \bibinfo{pages}{210602} (\bibinfo{year}{2019}).

\bibitem[{\citenamefont{Khasseh et~al.}(2019)\citenamefont{Khasseh, Fazio,
  Ruffo, and Russomanno}}]{Khasseh2019}
\bibinfo{author}{\bibfnamefont{R.}~\bibnamefont{Khasseh}},
  \bibinfo{author}{\bibfnamefont{R.}~\bibnamefont{Fazio}},
  \bibinfo{author}{\bibfnamefont{S.}~\bibnamefont{Ruffo}}, \bibnamefont{and}
  \bibinfo{author}{\bibfnamefont{A.}~\bibnamefont{Russomanno}},
  \bibinfo{journal}{Phys. Rev. Lett.} \textbf{\bibinfo{volume}{123}},
  \bibinfo{pages}{184301} (\bibinfo{year}{2019}).

\bibitem[{\citenamefont{Cai et~al.}(2020)\citenamefont{Cai, Huang, and
  Liu}}]{Cai2020}
\bibinfo{author}{\bibfnamefont{Z.}~\bibnamefont{Cai}},
  \bibinfo{author}{\bibfnamefont{Y.}~\bibnamefont{Huang}}, \bibnamefont{and}
  \bibinfo{author}{\bibfnamefont{W.~V.} \bibnamefont{Liu}},
  \bibinfo{journal}{Chinese Physics Letters} \textbf{\bibinfo{volume}{37}},
  \bibinfo{pages}{050503} (\bibinfo{year}{2020}).

\bibitem[{\citenamefont{{Chinzei} and {Ikeda}}(2020)}]{Chinzei2020}
\bibinfo{author}{\bibfnamefont{K.}~\bibnamefont{{Chinzei}}} \bibnamefont{and}
  \bibinfo{author}{\bibfnamefont{T.~N.} \bibnamefont{{Ikeda}}},
  \bibinfo{journal}{arXiv e-prints} \bibinfo{eid}{arXiv:2003.13315}
  (\bibinfo{year}{2020}), \eprint{2003.13315}.

\bibitem[{\citenamefont{Yao et~al.}(2020)\citenamefont{Yao, Nayak, Balents, and
  Zaletel}}]{Yao2020}
\bibinfo{author}{\bibfnamefont{N.~Y.} \bibnamefont{Yao}},
  \bibinfo{author}{\bibfnamefont{C.}~\bibnamefont{Nayak}},
  \bibinfo{author}{\bibfnamefont{L.}~\bibnamefont{Balents}}, \bibnamefont{and}
  \bibinfo{author}{\bibfnamefont{M.~P.} \bibnamefont{Zaletel}},
  \bibinfo{journal}{Nat. Phys.} \textbf{\bibinfo{volume}{16}},
  \bibinfo{pages}{438} (\bibinfo{year}{2020}).

\bibitem[{\citenamefont{Bruno}(2013)}]{Bruno2013}
\bibinfo{author}{\bibfnamefont{P.}~\bibnamefont{Bruno}},
  \bibinfo{journal}{Phys. Rev. Lett.} \textbf{\bibinfo{volume}{111}},
  \bibinfo{pages}{070402} (\bibinfo{year}{2013}).

\bibitem[{\citenamefont{Watanabe and Oshikawa}(2015)}]{Watanabe2015}
\bibinfo{author}{\bibfnamefont{H.}~\bibnamefont{Watanabe}} \bibnamefont{and}
  \bibinfo{author}{\bibfnamefont{M.}~\bibnamefont{Oshikawa}},
  \bibinfo{journal}{Phys. Rev. Lett.} \textbf{\bibinfo{volume}{114}},
  \bibinfo{pages}{251603} (\bibinfo{year}{2015}).

\bibitem[{\citenamefont{Choi et~al.}(2017)\citenamefont{Choi, Landig, Kucsko,
  Zhou, Isoya, Jelezko, Onoda, Sumiya, Khemani, von Keyserlingk
  et~al.}}]{Choi2017}
\bibinfo{author}{\bibfnamefont{S.}~\bibnamefont{Choi}},
  \bibinfo{author}{\bibfnamefont{R.}~\bibnamefont{Landig}},
  \bibinfo{author}{\bibfnamefont{G.}~\bibnamefont{Kucsko}},
  \bibinfo{author}{\bibfnamefont{H.}~\bibnamefont{Zhou}},
  \bibinfo{author}{\bibfnamefont{J.}~\bibnamefont{Isoya}},
  \bibinfo{author}{\bibfnamefont{F.}~\bibnamefont{Jelezko}},
  \bibinfo{author}{\bibfnamefont{S.}~\bibnamefont{Onoda}},
  \bibinfo{author}{\bibfnamefont{H.}~\bibnamefont{Sumiya}},
  \bibinfo{author}{\bibfnamefont{V.}~\bibnamefont{Khemani}},
  \bibinfo{author}{\bibfnamefont{C.}~\bibnamefont{von Keyserlingk}},
  \bibnamefont{et~al.}, \bibinfo{journal}{Nature}
  \textbf{\bibinfo{volume}{543}}, \bibinfo{pages}{221} (\bibinfo{year}{2017}).

\bibitem[{\citenamefont{Zhang et~al.}(2017)\citenamefont{Zhang, Hess,
  Kyprianidis, Becker, Lee, Smith, Pagano, Potirniche, Potter, Vishwanath
  et~al.}}]{Zhang2017}
\bibinfo{author}{\bibfnamefont{J.}~\bibnamefont{Zhang}},
  \bibinfo{author}{\bibfnamefont{P.~W.} \bibnamefont{Hess}},
  \bibinfo{author}{\bibfnamefont{A.}~\bibnamefont{Kyprianidis}},
  \bibinfo{author}{\bibfnamefont{P.}~\bibnamefont{Becker}},
  \bibinfo{author}{\bibfnamefont{A.}~\bibnamefont{Lee}},
  \bibinfo{author}{\bibfnamefont{J.}~\bibnamefont{Smith}},
  \bibinfo{author}{\bibfnamefont{G.}~\bibnamefont{Pagano}},
  \bibinfo{author}{\bibfnamefont{I.-D.} \bibnamefont{Potirniche}},
  \bibinfo{author}{\bibfnamefont{A.~C.} \bibnamefont{Potter}},
  \bibinfo{author}{\bibfnamefont{A.}~\bibnamefont{Vishwanath}},
  \bibnamefont{et~al.}, \bibinfo{journal}{Nature}
  \textbf{\bibinfo{volume}{543}}, \bibinfo{pages}{217} (\bibinfo{year}{2017}).

\bibitem[{\citenamefont{Su et~al.}(1979)\citenamefont{Su, Schrieffer, and
  Heeger}}]{Su1979}
\bibinfo{author}{\bibfnamefont{W.~P.} \bibnamefont{Su}},
  \bibinfo{author}{\bibfnamefont{J.~R.} \bibnamefont{Schrieffer}},
  \bibnamefont{and} \bibinfo{author}{\bibfnamefont{A.~J.}
  \bibnamefont{Heeger}}, \bibinfo{journal}{Phys. Rev. Lett.}
  \textbf{\bibinfo{volume}{42}}, \bibinfo{pages}{1698} (\bibinfo{year}{1979}).

\bibitem[{\citenamefont{Hohenberg and Halperin}(1977)}]{Hohenberg1977}
\bibinfo{author}{\bibfnamefont{P.~C.} \bibnamefont{Hohenberg}}
  \bibnamefont{and} \bibinfo{author}{\bibfnamefont{B.~I.}
  \bibnamefont{Halperin}}, \bibinfo{journal}{Rev. Mod. Phys.}
  \textbf{\bibinfo{volume}{49}}, \bibinfo{pages}{435} (\bibinfo{year}{1977}).

\bibitem[{\citenamefont{Taeuber}(2014)}]{Taeuber2014}
\bibinfo{author}{\bibfnamefont{U.}~\bibnamefont{Taeuber}},
  \emph{\bibinfo{title}{CRITICAL DYNAMICS: A Field Theory Approach to
  Equilibrium and Non-Equilibrium Scaling Behavior}}
  (\bibinfo{publisher}{~Cambridge University Press, Cambridge},
  \bibinfo{year}{2014}).

\bibitem[{\citenamefont{Landig et~al.}(2016)\citenamefont{Landig, Hruby, Dogra,
  Landini, R.Mottl, Donner, and Esslinger}}]{Landig2016}
\bibinfo{author}{\bibfnamefont{R.}~\bibnamefont{Landig}},
  \bibinfo{author}{\bibfnamefont{L.}~\bibnamefont{Hruby}},
  \bibinfo{author}{\bibfnamefont{N.}~\bibnamefont{Dogra}},
  \bibinfo{author}{\bibfnamefont{M.}~\bibnamefont{Landini}},
  \bibinfo{author}{\bibnamefont{R.Mottl}},
  \bibinfo{author}{\bibfnamefont{T.}~\bibnamefont{Donner}}, \bibnamefont{and}
  \bibinfo{author}{\bibfnamefont{T.}~\bibnamefont{Esslinger}},
  \bibinfo{journal}{Nature} \textbf{\bibinfo{volume}{532}},
  \bibinfo{pages}{476} (\bibinfo{year}{2016}).

\bibitem[{\citenamefont{Hruby et~al.}(2018)\citenamefont{Hruby, Dogra, Landini,
  Donner, and Esslinger}}]{Hruby2018}
\bibinfo{author}{\bibfnamefont{L.}~\bibnamefont{Hruby}},
  \bibinfo{author}{\bibfnamefont{N.}~\bibnamefont{Dogra}},
  \bibinfo{author}{\bibfnamefont{M.}~\bibnamefont{Landini}},
  \bibinfo{author}{\bibfnamefont{T.}~\bibnamefont{Donner}}, \bibnamefont{and}
  \bibinfo{author}{\bibfnamefont{T.}~\bibnamefont{Esslinger}},
  \bibinfo{journal}{PNAS} \textbf{\bibinfo{volume}{115}}, \bibinfo{pages}{3279}
  (\bibinfo{year}{2018}).

\bibitem[{\citenamefont{Bla\ss{} et~al.}(2018)\citenamefont{Bla\ss{}, Rieger,
  Ro\'osz, and Igl\'oi}}]{Blass2018}
\bibinfo{author}{\bibfnamefont{B.}~\bibnamefont{Bla\ss{}}},
  \bibinfo{author}{\bibfnamefont{H.}~\bibnamefont{Rieger}},
  \bibinfo{author}{\bibfnamefont{G.~m.~H.} \bibnamefont{Ro\'osz}},
  \bibnamefont{and} \bibinfo{author}{\bibfnamefont{F.}~\bibnamefont{Igl\'oi}},
  \bibinfo{journal}{Phys. Rev. Lett.} \textbf{\bibinfo{volume}{121}},
  \bibinfo{pages}{095301} (\bibinfo{year}{2018}).

\bibitem[{\citenamefont{Igl\'oi et~al.}(2018)\citenamefont{Igl\'oi, Bla\ss{},
  Ro\'osz, and Rieger}}]{Igloi2018}
\bibinfo{author}{\bibfnamefont{F.}~\bibnamefont{Igl\'oi}},
  \bibinfo{author}{\bibfnamefont{B.}~\bibnamefont{Bla\ss{}}},
  \bibinfo{author}{\bibfnamefont{G.~m.~H.} \bibnamefont{Ro\'osz}},
  \bibnamefont{and} \bibinfo{author}{\bibfnamefont{H.}~\bibnamefont{Rieger}},
  \bibinfo{journal}{Phys. Rev. B} \textbf{\bibinfo{volume}{98}},
  \bibinfo{pages}{184415} (\bibinfo{year}{2018}).

\bibitem[{Sup()}]{Supplementary}
\bibinfo{howpublished}{See the Supplementary Material for the details of the
  mean-field method, the analysis of the quantum quench and periodically driven
  dynamics and the convergence check of our numerical results}.

\bibitem[{\citenamefont{Chen and Cai}(2020)}]{Chen2020}
\bibinfo{author}{\bibfnamefont{Y.}~\bibnamefont{Chen}} \bibnamefont{and}
  \bibinfo{author}{\bibfnamefont{Z.}~\bibnamefont{Cai}},
  \bibinfo{journal}{Phys. Rev. A} \textbf{\bibinfo{volume}{101}},
  \bibinfo{pages}{023611} (\bibinfo{year}{2020}).

\bibitem[{\citenamefont{Russomanno et~al.}(2012)\citenamefont{Russomanno,
  Silva, and Santoro}}]{Russomanno2012}
\bibinfo{author}{\bibfnamefont{A.}~\bibnamefont{Russomanno}},
  \bibinfo{author}{\bibfnamefont{A.}~\bibnamefont{Silva}}, \bibnamefont{and}
  \bibinfo{author}{\bibfnamefont{G.~E.} \bibnamefont{Santoro}},
  \bibinfo{journal}{Phys. Rev. Lett.} \textbf{\bibinfo{volume}{109}},
  \bibinfo{pages}{257201} (\bibinfo{year}{2012}).

\bibitem[{\citenamefont{{Else} et~al.}(2019)\citenamefont{{Else}, {Monroe},
  {Nayak}, and {Yao}}}]{Else2019}
\bibinfo{author}{\bibfnamefont{D.~V.} \bibnamefont{{Else}}},
  \bibinfo{author}{\bibfnamefont{C.}~\bibnamefont{{Monroe}}},
  \bibinfo{author}{\bibfnamefont{C.}~\bibnamefont{{Nayak}}}, \bibnamefont{and}
  \bibinfo{author}{\bibfnamefont{N.~Y.} \bibnamefont{{Yao}}},
  \bibinfo{journal}{arXiv e-prints} \bibinfo{eid}{arXiv:1905.13232}
  (\bibinfo{year}{2019}), \eprint{1905.13232}.

\bibitem[{\citenamefont{J.M.T.Thompson and H.B.Stewart}(2002)}]{Thompson2002}
\bibinfo{author}{\bibnamefont{J.M.T.Thompson}} \bibnamefont{and}
  \bibinfo{author}{\bibnamefont{H.B.Stewart}}, \emph{\bibinfo{title}{Nonlinear
  Dynamcis and Chaos}} (\bibinfo{publisher}{John Wiley and Sons, LTD},
  \bibinfo{year}{2002}).

\bibitem[{\citenamefont{Yuzbashyan et~al.}(2005)\citenamefont{Yuzbashyan,
  Altshuler, Kuznetsov, and Enolskii}}]{Yuzbashyan2005}
\bibinfo{author}{\bibfnamefont{E.~A.} \bibnamefont{Yuzbashyan}},
  \bibinfo{author}{\bibfnamefont{B.~L.} \bibnamefont{Altshuler}},
  \bibinfo{author}{\bibfnamefont{V.~B.} \bibnamefont{Kuznetsov}},
  \bibnamefont{and} \bibinfo{author}{\bibfnamefont{V.~Z.}
  \bibnamefont{Enolskii}}, \bibinfo{journal}{Phys. Rev. B}
  \textbf{\bibinfo{volume}{72}}, \bibinfo{pages}{220503}
  (\bibinfo{year}{2005}).

\bibitem[{\citenamefont{Trotzky et~al.}(2012)\citenamefont{Trotzky, Chen,
  Flesch, McCulloch, Schollw\"{o}ck, Eisert, and Bloch}}]{Trotzky2012}
\bibinfo{author}{\bibfnamefont{S.}~\bibnamefont{Trotzky}},
  \bibinfo{author}{\bibfnamefont{Y.-A.} \bibnamefont{Chen}},
  \bibinfo{author}{\bibfnamefont{A.}~\bibnamefont{Flesch}},
  \bibinfo{author}{\bibfnamefont{I.~P.} \bibnamefont{McCulloch}},
  \bibinfo{author}{\bibfnamefont{U.}~\bibnamefont{Schollw\"{o}ck}},
  \bibinfo{author}{\bibfnamefont{J.}~\bibnamefont{Eisert}}, \bibnamefont{and}
  \bibinfo{author}{\bibfnamefont{I.}~\bibnamefont{Bloch}},
  \bibinfo{journal}{Nature Phys} \textbf{\bibinfo{volume}{8}},
  \bibinfo{pages}{325} (\bibinfo{year}{2012}).

\bibitem[{\citenamefont{Autti et~al.}(2018)\citenamefont{Autti, Eltsov, and
  Volovik}}]{Autti2018}
\bibinfo{author}{\bibfnamefont{S.}~\bibnamefont{Autti}},
  \bibinfo{author}{\bibfnamefont{V.~B.} \bibnamefont{Eltsov}},
  \bibnamefont{and} \bibinfo{author}{\bibfnamefont{G.~E.}
  \bibnamefont{Volovik}}, \bibinfo{journal}{Phys. Rev. Lett.}
  \textbf{\bibinfo{volume}{120}}, \bibinfo{pages}{215301}
  (\bibinfo{year}{2018}).

\bibitem[{\citenamefont{{Hayata} and {Hidaka}}(2018)}]{Hayata2018}
\bibinfo{author}{\bibfnamefont{T.}~\bibnamefont{{Hayata}}} \bibnamefont{and}
  \bibinfo{author}{\bibfnamefont{Y.}~\bibnamefont{{Hidaka}}},
  \bibinfo{journal}{arXiv e-prints} \bibinfo{eid}{arXiv:1808.07636}
  (\bibinfo{year}{2018}), \eprint{1808.07636}.

\end{thebibliography}

\end{document}